\documentclass[%
 reprint,
 showpacs,
 amsmath,amssymb,
 aps,
 pra,
]{revtex4-1}

\usepackage[ascii]{inputenc}
\usepackage[T1]{fontenc}
\usepackage[english]{babel}
\usepackage{graphicx}
\usepackage{microtype}
\usepackage[pdftex,colorlinks]{hyperref}
\hypersetup{%
  linkcolor=blue,%
  citecolor=blue,%
  urlcolor=blue,%
  pdftitle={Cusp bifurcation in the eigenvalue spectrum of PT-symmetric
    Bose-Einstein condensates},%
  pdfauthor={Daniel Dizdarevic, Dennis Dast, Daniel Haag, J\"org Main,
    Holger Cartarius, G\"unter Wunner}%
}

\DeclareMathOperator{\imag}{Im}
\DeclareMathOperator{\real}{Re}
\DeclareMathOperator{\trace}{tr}
\DeclareMathOperator{\diag}{diag}
\newcommand{\PT}{\mathcal{PT}}
\newcommand{\mbf}{\mathbf}
\newcommand{\mrm}{\mathrm}
\newcommand{\rmi}{\mathrm{i}}
\newcommand{\rmj}{\mathrm{j}}
\newcommand{\rmk}{\mathrm{k}}

\newcommand{\ket}[1]{|{#1}\rangle}
\newcommand{\braket}[2]{\langle{#1}|{#2}\rangle}

\begin{document}

\title{Cusp bifurcation in the eigenvalue spectrum of $\PT$-symmetric
Bose-Einstein condensates}

\author{Daniel Dizdarevic}

\author{Dennis Dast}
\email[]{dennis.dast@itp1.uni-stuttgart.de}

\author{Daniel Haag}

\author{J\"org Main}

\author{Holger Cartarius}

\author{G\"unter Wunner}

\affiliation{Institut f\"ur Theoretische Physik 1,
  Universit\"at Stuttgart, 70550 Stuttgart, Germany}

\date{\today}

\begin{abstract}
  A Bose-Einstein condensate in a double-well potential features stationary
  solutions even for attractive contact interaction as long as the particle
  number and therefore the interaction strength do not exceed a certain limit.
  Introducing balanced gain and loss into such a system drastically changes the
  bifurcation scenario at which these states are created.
  Instead of two tangent bifurcations at which the symmetric and
  antisymmetric states emerge, one tangent bifurcation between two formerly
  independent branches arises [D. Haag \textit{et al.},
  \href{http://dx.doi.org/10.1103/PhysRevA.89.023601}
  {Phys. Rev. A \textbf{89}, 023601 (2014)}].
  We study this transition in detail using a bicomplex formulation of the
  time-dependent variational principle and find that in fact there are three
  tangent bifurcations for very small gain-loss contributions which coalesce in
  a cusp bifurcation.
\end{abstract}


\maketitle

\section{Introduction}
\label{sec:introduction}
Bose-Einstein condensates with attractive contact interactions become unstable
if the number of particles exceeds a certain limit \cite{Ruprecht95a, Dodd96a,
Houbiers96a}.
In mean-field approximation where the condensate is described by the
Gross-Pitaevskii equation this effect manifests itself in a vanishing ground
state.
Above this limit the mean interaction caused by the particles is strong enough
to constrict and ultimately collapse the condensate.
Below this limit the stationary solutions of the Gross-Pitaevskii equation are
observable even though for attractive interactions the ground state is not the
global minimum of the mean-field energy \cite{Sackett97a}.

Characteristic properties of Bose-Einstein condensates with attractive
interaction could be used for an atomic soliton laser \cite{Ruprecht95a}.
However, this requires the realization of a particle flow into and out of the
condensate.
Modeling such effects in mean-field approximation is done via imaginary
potentials, thus rendering the Hamiltonian non-Hermitian \cite{Moiseyev11a,
Kagan98a}.
Such non-Hermitian systems have been widely studied \cite{Kagan98a,
Schlagheck06a, Rapedius09a, Rapedius10a, Abdullaev10a, Bludov10a, Witthaut11a}
and are supported by comparison with many-particle calculations
\cite{Rapedius13a, Dast14a}.
Both in- and outcoupling of particles have been experimentally realized
\cite{Gericke08a, Robins08a}.

Despite the non-Hermiticity real eigenvalues and thus true stationary states
can be found if the Hamiltonian is $\PT$-symmetric \cite{Bender98a, Bender99a,
Bender07a} or pseudo-Hermitian \cite{Mostafazadeh02a, Mostafazadeh02b,
Mostafazadeh02c}.
This unique property motivated a variety of theoretical studies
\cite{Klaiman08a, Schindler11a, Bittner12a, Cartarius12a, Cartarius12b,
Mayteevarunyoo13a, Graefe08a, Graefe08b, Graefe12a} and the experimental
realization in optical wave-guide systems \cite{Klaiman08a, Ruter10a, Guo09a,
Peng14b}.

In \cite{Haag14a} we studied the eigenvalue spectrum and dynamical properties
of a Bose-Einstein condensate, described by the dimensionless Gross-Pitaevskii
equation
\begin{equation}
  (-\Delta + V_\mathrm{ext} + 8 \pi Na |\psi|^2) \psi = \mu \psi,
  \label{eq:GPE}
\end{equation}
in a three-dimensional $\PT$-symmetric double-well potential
\begin{equation}
  V_\mathrm{ext}(\mbf{x}) = \frac14 [x^2 + \omega^2(y^2+z^2)] + v_0
    \mrm{e}^{-\sigma x^2} + \rmi \gamma x \mrm{e}^{-\rho x^2},
  \label{eq:potential}
\end{equation}
with $\omega = 2$, $v_0 = 4$, $\sigma = 0.5$ and $\rho \approx 0.12$.
The symmetric real part of the potential is a harmonic trap which is separated
into two wells by a Gaussian barrier.
The antisymmetric imaginary part of the potential induces particle loss in the
left well and particle gain in the right well whose strength is given by the
gain-loss parameter $\gamma$.

It was found that four stationary states of the real double-well
potential are created in the two lowest-lying tangent bifurcations at strong
attractive interaction strengths $Na$.
If gain and loss is introduced into the system, instead of two bifurcations
there is only one bifurcation between two previously independent states, while
the other two states vanish.
However, the underlying bifurcation mechanism remained unclear.
It is the purpose of this paper to clarify this mechanism and study the
bifurcation scenario in detail classifying it as a cusp bifurcation
\cite{Poston78a}.

The number of solutions is not conserved at bifurcations since the
Gross-Pitaevskii equation is non-analytic due to its nonlinear part $|\psi|^2$.
This problem is addressed by applying an analytic continuation, where the
complex wave functions are replaced by bicomplex ones.
The numerical results are then obtained via the \textit{time-dependent
variational principle} (TDVP).

We start with a short introduction to bicomplex numbers in
Sec.~\ref{sec:bicomplexnumbers} before discussing their application to the
Gross-Pitaevskii equation and the TDVP in Sec.~\ref{sec:numericalmethod}.
A detailed analysis of the eigenvalue spectrum and the bifurcation scenario is
carried out in Sec.~\ref{sec:results}.
Conclusions are drawn in Sec.~\ref{sec:conclusion}.

\section{Bicomplex numbers}
\label{sec:bicomplexnumbers}
Consider a usual complex number $z \in \mathbb{C}$, $z= x + \rmi y$ with $x,y
\in \mathbb{R}$ and the imaginary unit $\rmi^2=-1$.
A bicomplex number is constructed by replacing the real and imaginary part of
$z$ by complex numbers with an additional imaginary unit $\rmj$ which also
fulfills $\rmj^2=-1$,
\begin{equation}
  z = x + \rmi y = (x_1 + \rmj x_\rmj) + \rmi (y_1 + \rmj y_\rmj) \equiv
    z_1 + \rmi z_\rmi + \rmj z_\rmj + \rmk z_\rmk.
  \label{eq:bicomplex_components}
\end{equation}
In the last step $\rmk=\rmi\rmj$ with $\rmk^2=1$ was introduced.

Every bicomplex number can be uniquely written \cite{GervaisLavoie10a,
GervaisLavoie11a} as
\begin{equation}
  z= z_+ e_+ + z_- e_-,
  \label{eq:bicomplex_idempotent}
\end{equation}
with the components
\begin{subequations}
  \begin{align}
    z_+ &= (z_1 + z_\rmk) + \rmi (z_\rmi - z_\rmj),\\
    z_- &= (z_1 - z_\rmk) + \rmi (z_\rmi + z_\rmj),
  \end{align}
  \label{eq:pm_components}%
\end{subequations}
and the idempotent basis
\begin{equation}
  e_\pm = \frac{1\pm\rmk}{2}.
  \label{eq:idempotent_basis}
\end{equation}
One can easily confirm that $e_\pm$ fulfill the relations
\begin{equation}
  e_+ + e_- = 1,\quad e_+ - e_- = \rmk, \quad e_+e_-=0, \quad e_\pm^2=e_\pm.
  \label{eq:basis_properties}
\end{equation}
Note that an equivalent representation exists in which the components $z_\pm$
are complex numbers with the imaginary unit $\rmj$ instead of the imaginary
unit $\rmi$.

The introduction of the idempotent basis significantly simplifies arithmetic
operations of bicomplex numbers since for two bicomplex numbers $z, w$ the
relation
\begin{equation}
  z \circ w = (z_+ \circ w_+)e_+ + (z_- \circ w_-) e_-
  \label{eq:bc_arithmetic}
\end{equation}
holds, where $\circ$ represents addition, subtraction, multiplication and
division.
Thus the bicomplex algebra is effectively reduced to a complex algebra within
the coefficients.

\section{Numerical method}
\label{sec:numericalmethod}
As a next step we apply the analytic continuation to the Gross-Pitaevskii
equation~\eqref{eq:GPE} by allowing bicomplex values for the wave function
$\psi$ and the chemical potential $\mu$.
With the notation introduced in Eq.~\eqref{eq:bicomplex_idempotent} the modulus
squared can be written as $|\psi|^2 = \psi \psi^* = \psi_+ \psi_-^* e_+ +
\psi_- \psi_+^* e_-$.
Here $\psi^*$ denotes a complex conjugation of $\psi$ with respect to the
complex unit $\rmi$.
Since $\rmk=\rmi\rmj$ this complex conjugation changes also the sign of $\rmk$
and thus transforms $e_+$ into $e_-$ and vice versa.
Due to the properties \eqref{eq:basis_properties} and \eqref{eq:bc_arithmetic}
the $\pm$ components are independent and the bicomplex Gross-Pitaevskii
equation can be split into two coupled complex differential equations, which
evidently do not contain a non-analytical term anymore,
\begin{subequations}
  \begin{align}
    (-\Delta + V_\mathrm{ext} + 8\pi Na \psi_+ \psi_-^*) \psi_+ &= \mu_+
      \psi_+, \\
    (-\Delta - V_\mathrm{ext} + 8\pi Na \psi_- \psi_+^*) \psi_- &= \mu_-
      \psi_-.
  \end{align}
  \label{eq:coupled_GPE}%
\end{subequations}
For complex wave functions the relation $\psi_+=\psi_-$ holds, as can be seen
in Eqs.~\eqref{eq:pm_components}, and the Eqs.~\eqref{eq:coupled_GPE} are
reduced to the Gross-Pitaevskii equation \eqref{eq:GPE}.
Thus, the coupled Eqs.~\eqref{eq:coupled_GPE} still support all solutions of
the Gross-Pitaevskii equation \eqref{eq:GPE}.

To solve the Eqs.~\eqref{eq:coupled_GPE} the TDVP is generalized for bicomplex
differential equations.
Our ansatz consists of coupled Gaussian wave packets which yields accurate
results even for a small number of wave packets \cite{Rau10b, Rau10c, Rau10a}.
For the double-well potential studied in this work it was shown that two
coupled Gaussian wave packets suffice and additional wave packets only lead to
small corrections \cite{Dast13a, Haag14a}.
For all calculations the following bicomplex ansatz is used
\begin{equation}
  \psi(\mbf{x}, \mbf{z}(t)) = \sum_{n=1}^{2} \exp(-\mbf{x}^T A^n \mbf{x}
  + \mbf{b}^n \cdot \mbf{x} + c^n),
  \label{eq:ansatz}
\end{equation}
with the bicomplex three-dimensional diagonal matrix $A^n=\diag(A^n_\parallel,
A^n_\perp, A^n_\perp)$, the bicomplex three-dimensional vector $\mbf{b}^n=(b^n,
0, 0)^T$, and the bicomplex number $c^n$.
The vector $\mbf{z}(t)$ contains all parameters $A^n,$ $\mbf{b}^n$, $c^n$ of
the wave function.
The ansatz is chosen such that all solutions are invariant under rotations
around the $x$ axis, i.e.\ we search only for wave functions that possess the
symmetry of the potential.
The complex $\pm$ components of the wave function can be written as
\begin{align}
  \psi_\pm(\mbf{x}, \mbf{z}_\pm(t)) &= \sum_{n=1}^{2} \exp(-\mbf{x}^T A_\pm^n
  \mbf{x} + \mbf{b}_\pm^n \cdot \mbf{x} + c_\pm^n) \notag \\
  & \equiv \sum_{n=1}^2 g_\pm^n
  \label{eq:ansatz_pm}
\end{align}
using Eq.~\eqref{eq:bc_arithmetic} since the exponential function is defined
as a power series.

The TDVP makes use of the McLachlan variational principle \cite{McLachlan64a}
which demands that the variation of the functional
\begin{equation}
  I = \|\rmi \varphi - \mathcal{H}\psi \|^2
  \label{eq:functional}
\end{equation}
with respect to $\varphi$ vanishes, then after the variation
$\varphi=\dot{\psi}$ is set.
Using the representation with the idempotent basis
\eqref{eq:bicomplex_idempotent} for $\varphi$, $\psi$ and $\mathcal{H}$ leads
to
\begin{align}
  I = &\braket{\rmi \varphi_- - \mathcal{H}_- \psi_-}{\rmi \varphi_+ -
    \mathcal{H}_+ \psi_+} e_+ \notag\\
    & + \braket{\rmi \varphi_+ - \mathcal{H}_+ \psi_+}{\rmi \varphi_- -
    \mathcal{H}_- \psi_-} e_- \notag\\
    \equiv & I_+ e_+ + I_- e_-
  \label{eq:functional_idempotent}
\end{align}
The two coefficients are complex conjugate, thus, the functional
$\tilde{I}=I_+=I_-^*$ already contains the full information.
The variation of the functional $\tilde{I}$ reads,
\begin{align}
  \delta\tilde{I} &= \braket{\rmi \delta \varphi_-}{\rmi \varphi_+ -
    \mathcal{H}_+ \psi_+}
    + \braket{\rmi \varphi_- - \mathcal{H}_- \psi_-}{\rmi \delta \varphi_+}
    \notag \\
  &= \braket{\rmi \partial_{\mbf{z}_-} \psi_-}{\rmi \dot{\psi}_+ -
    \mathcal{H}_+ \psi_+} \delta \dot{\mbf{z}}_-^* \notag\\
  & \quad + \braket{\rmi \dot{\psi}_- - \mathcal{H}_- \psi_-}{\rmi
    \partial_{\mbf{z}_+} \psi_+} \delta \dot{\mbf{z}}_+ \notag\\
  & = 0.
  \label{eq:functional_variation}
\end{align}
Since the variations $\delta \dot{\mbf{z}}_+$ and $\delta \dot{\mbf{z}}_-^*$
are independent both coefficients have to vanish, which can be written in the
compact form
\begin{equation}
  \braket{\rmi\dot{\psi}_\mp - \mathcal{H}_\mp \psi_\mp}
    {\partial_{\mbf{z}_\pm} \psi_\pm} = 0.
  \label{eq:functional_variation_compact}
\end{equation}
Using the idempotent basis we can reduce the bicomplex equation to a pair of
coupled complex equations.

The next step is to insert the ansatz \eqref{eq:ansatz_pm} of the wave
function into Eq.~\eqref{eq:functional_variation_compact}.
The calculations are in full analogy with the TDVP for complex Gaussian wave
packets \cite{Eichler12a} and therefore we only list the results.

The equations of motion for the parameters of the Gaussian wave packets read
\begin{subequations}
  \begin{align}
    \rmi \dot{A}_\pm^n &= 4 (A_\pm^n)^2 - V_{2\pm}^n, \\
    \rmi \dot{\mbf{b}}_\pm^n &= 4 A_\pm^n \mbf{b}_\pm^n + \mbf{v}_{1\pm}^n, \\
    \rmi \dot{c}_\pm^n &= 2 \trace{A}_\pm^n - \mbf{b}_\pm^n \cdot \mbf{b}_\pm^n
      + v_{0\pm}^n.
  \end{align}
  \label{eq:eom}%
\end{subequations}
The coefficients of the effective potential $V_{2\pm}^n$, $\mbf{v}_{1\pm}^n$
and $v_{0\pm}$ are obtained by numerically solving
\begin{align}
  \sum_{n=1}^N \braket{x_\alpha^k x_\beta^l g_\mp^m}
    {(\mbf{x}^T V_{2\pm}^n \mbf{x} + \mbf{v}_{1\pm}^n \cdot \mbf{x} +
    v_{0\pm}^n) g_\pm^n} \notag\\
  = \sum_{n=1}^N \braket{x_\alpha^k x_\beta^l g_\mp^m}
  {(V_\mrm{ext} + 8\pi Na \psi_\mp^* \psi_\pm) g_\pm^n},
  \label{eq:effective_potential}
\end{align}
with $\alpha, \beta = 1,\dots,d$; $m=1,\dots,N$ and $k+l=0,1,2$,
where $d$ is the dimension  and $N$ the number of coupled Gaussians.
For the ansatz \eqref{eq:ansatz_pm} used in this work we have $d=3$ and $N=2$.

Stationary solutions are obtained by varying the parameters $A_\pm^n$,
$\mbf{b}_\pm^n$, $(c_\pm^n-c_\pm^N)$, such that they fulfill the equations
$\dot{A}_\pm^n=0$, $\dot{\mbf{b}}_\pm^n=0$, $(\dot{c}_\pm^n-\dot{c}_\pm^N)=0$
for $n=1,\dots,N$.
Note that the differences $(c_\pm^n-c_\pm^N)$ are used since not all
parameters $c_\pm^n$ are free due to the norm constraint and the choice of a
global phase.
Standard numerical methods for complex numbers can be used to achieve this
since all parameters in Eqs.~\eqref{eq:eom} and \eqref{eq:effective_potential}
are complex.
This again shows the benefit of the idempotent basis.

\section{Results}
\label{sec:results}
Before discussing the analytic continuation the real spectrum is investigated.
In \cite{Haag14a} the real eigenvalue spectrum of this system has already been
studied, however, the bifurcation scenario at strong attractive interactions,
on which we concentrate in this paper, was only discussed briefly.
In particular the bifurcation scenario was only shown for the gain-loss
parameters $\gamma=0$ and $\gamma=0.001$, which, as we will see, does not
suffice to understand the bifurcation process in detail.

\begin{figure}
  \centering
  \includegraphics[width=\columnwidth]{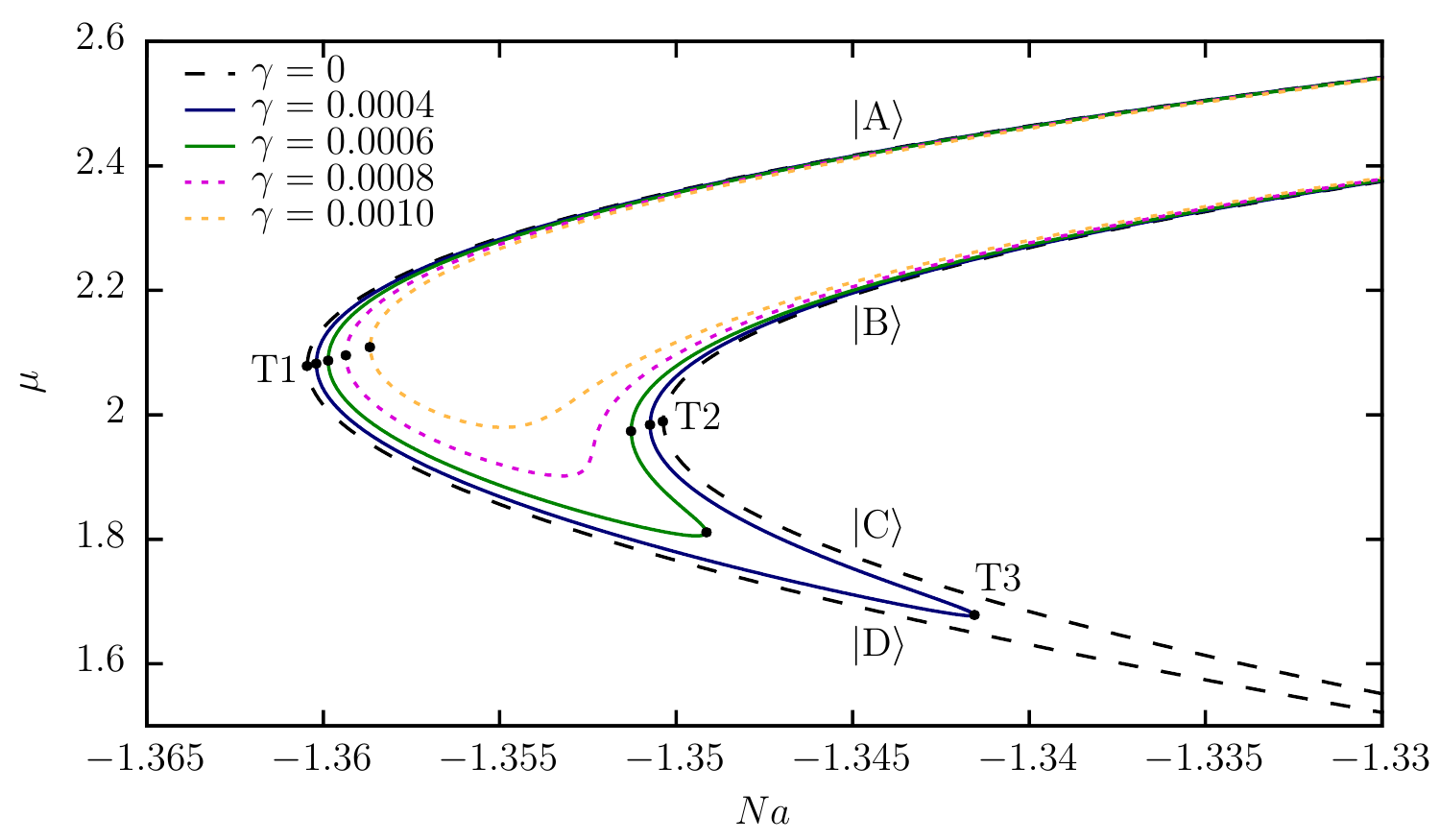}
  \caption{%
    (Color online)
    The bifurcation scenario of the $\PT$-symmetric stationary states of the
    three-dimensional double-well potential at strong attractive interactions.
    The chemical potential of the stationary solutions is shown as a function
    of the interaction strength $Na$ for different values of the gain-loss
    parameter $\gamma$.
    For $\gamma=0$ (black dashed line) two tangent bifurcations T1 and T2 are
    present.
    The states $\ket{\mrm{A}}$ and $\ket{\mrm{D}}$ emerge from the
    bifurcation T1 whereas T2 gives birth to the states $\ket{\mrm{B}}$ and
    $\ket{\mrm{C}}$.
    For small values of $\gamma$ (solid lines) a third bifurcation T3 appears
    at which the two states $\ket{\mrm{B}}$ and $\ket{\mrm{C}}$ bifurcate which
    connects the so far independent branches of the states $\ket{\mrm{C}}$ and
    $\ket{\mrm{D}}$.
    For higher values of $\gamma$ (dotted lines) the bifurcations T2 and T3
    vanish, as does the intermediate state $\ket{\mrm{C}}$, thus, the
    branches of the states $\ket{\mrm{B}}$ and $\ket{\mrm{D}}$ merge.
  }%
  \label{fig:real_spectrum}
\end{figure}

Figure~\ref{fig:real_spectrum}
shows the relevant part of the eigenvalue spectrum as a function of the
nonlinearity parameter $Na$ for different gain-loss parameters $\gamma$.
All states are $\PT$-symmetric, therefore, their chemical potentials $\mu$ are
real.
For a vanishing gain-loss contribution, $\gamma=0$, the two states
$\ket{\mrm{A}}$ and $\ket{\mrm{B}}$, which originate from the ground and
excited state of the double-well potential, emerge from independent tangent
bifurcations T1 and T2.
The bifurcation T1 gives birth to the states $\ket{\mrm{A}}$ and
$\ket{\mrm{D}}$ whereas the states $\ket{\mrm{B}}$ and $\ket{\mrm{C}}$ emerge
from the bifurcation T2.
At $\gamma=0.001$ the situation has changed fundamentally.
The two branches $\ket{\mrm{A}}$ and $\ket{\mrm{B}}$ are no longer independent
but emerge from the same bifurcation T1.
In addition the branches $\ket{\mrm{C}}$ and $\ket{\mrm{D}}$ have vanished.

This behavior can be understood by studying the system for very small
parameters $\gamma$.
For $\gamma=0.0004$ both tangent bifurcations T1 and T2, and thus, also the
branches $\ket{\mrm{C}}$ and $\ket{\mrm{D}}$ are still present.
However, the two pairs emerging from T1 and T2 are now connected by a new
tangent bifurcation T3 at which the lower lying states vanish.
If the gain-loss parameter is slightly increased the bifurcation T3
is shifted to lower values of $Na$ and approaches T2 ($\gamma=0.0006$).
For $\gamma=0.0008$ both bifurcations T2 and T3 have united and vanished.

\begin{figure}
  \centering
  \includegraphics[width=\columnwidth]{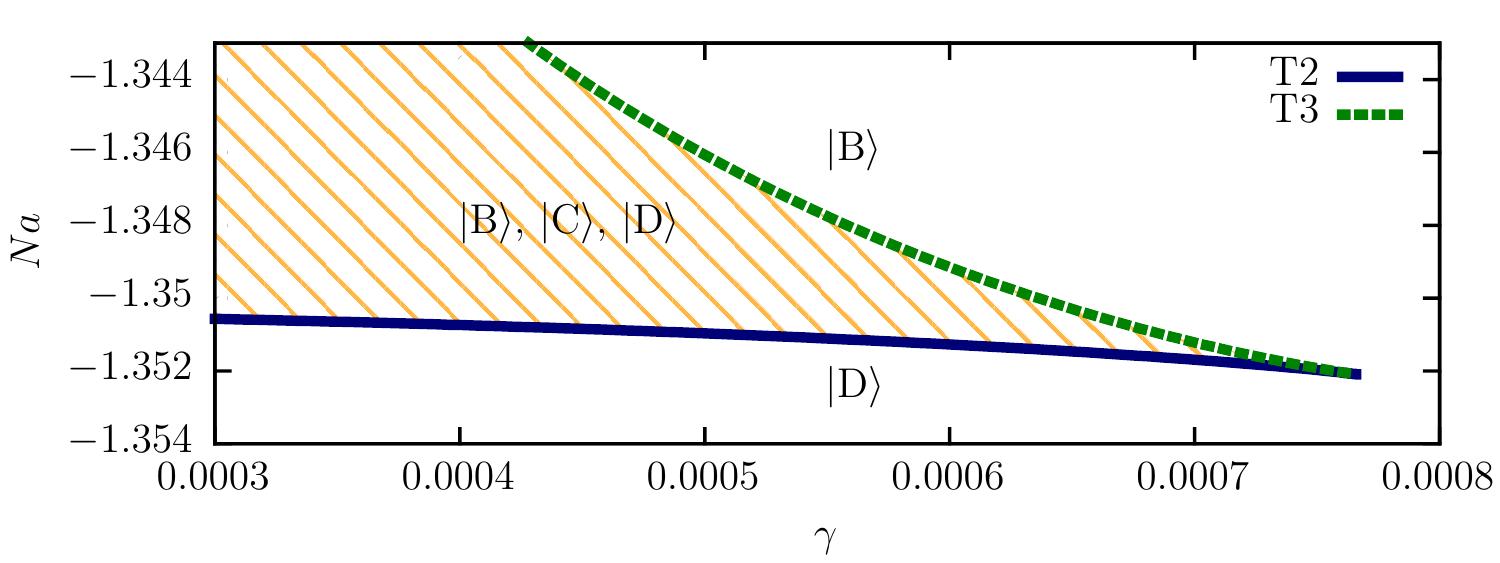}
  \caption{%
    (Color online)
    Trajectories of the two tangent bifurcations T2 and T3 in the
    $\gamma$-$Na$-parameter space.
    In the shaded area all three states involved in the two bifurcations are
    present while in the other areas only one of these states exists.
    For increasing values of the gain-loss parameter $\gamma$, the bifurcation
    T2 is strongly shifted to lower nonlinearity parameters $Na$.
    At $\gamma_\mrm{c} \approx 0.000766$ the two bifurcations coincide and the
    state $\ket{\mrm{C}}$ vanishes while the states $\ket{\mrm{B}}$ and
    $\ket{\mrm{D}}$ merge.
  }%
  \label{fig:phasediagram_spectrum}
\end{figure}

To discuss the propagation of the two tangent bifurcations in the parameter
space in more detail, a phase diagram is shown in
Fig.~\ref{fig:phasediagram_spectrum}.
The two bifurcations coalesce and vanish at critical values of the gain-loss
parameter $\gamma_\mrm{c} \approx 0.000766$ and interaction strength
$(Na)_\mrm{c} \approx -1.352$.
Their trajectories have the characteristic form of a cusp in the phase diagram.
In the area enclosed by the two trajectories all three states $\ket{\mrm{B}}$,
$\ket{\mrm{C}}$, and $\ket{\mrm{D}}$ involved in this bifurcation scenario are
present.
Outside of this area only one state remains.
Below the trajectory of T2 only the state $\ket{\mrm{D}}$ exists and above the
trajectory of T3 only the state $\ket{\mrm{B}}$.
Since the areas below T2 and above T3 are connected for $\gamma >
\gamma_\mrm{c}$, the states $\ket{\mrm{B}}$ and $\ket{\mrm{D}}$ can
continuously be transferred into each other by following a path in the
parameter space around the cusp.

We have seen that the transition from the $\gamma=0$ case, where the states
$\ket{\mrm{A}}$ and $\ket{\mrm{B}}$ are completely independent, to the
$\gamma=0.001$ case, where both states emerge from a common bifurcation,
occurs in two steps.
The first step is the formation of a new tangent bifurcation T3 between the
states $\ket{\mrm{C}}$ and $\ket{\mrm{D}}$ and the second step is the
coalescence of the two bifurcations.
In the following we will show that introducing bicomplex numbers to
analytically continue the Gross-Pitaevskii equation allows a more detailed
discussion of these two steps.

Bifurcation scenarios are classified by comparison with normal forms
\cite{Poston78a}.
The eigenvalue spectrum shows three tangent bifurcations which are described by
the normal form $\dot{x}=x^2 - \sigma$.
The stationary solutions, $\dot{x} = 0$, are obviously $x = \pm \sqrt{\sigma}$.
Two real solutions exist for $\sigma > 0$, they coalesce at $\sigma = 0$, where
two complex conjugate solutions ($\sigma < 0$) emerge.
Thus, the total number of solutions stays constant if we allow $x$ to be
complex.
Due to the non-analytic modulus squared in the Gross-Pitaevskii equation this
conservation of solutions is not guaranteed although the wave functions can
be complex.

In analogy with the transition from real to complex numbers in the normal form
of the tangent bifurcation, we expect a transition from complex to bicomplex
numbers in the eigenvalue spectrum of the Gross-Pitaevskii equation.
By introducing bicomplex numbers the Gross-Pitaevskii equation is analytically
continued and the two coupled analytic Eqs.~\eqref{eq:coupled_GPE} are
obtained.

\begin{figure}
  \centering
  \includegraphics[width=\columnwidth]{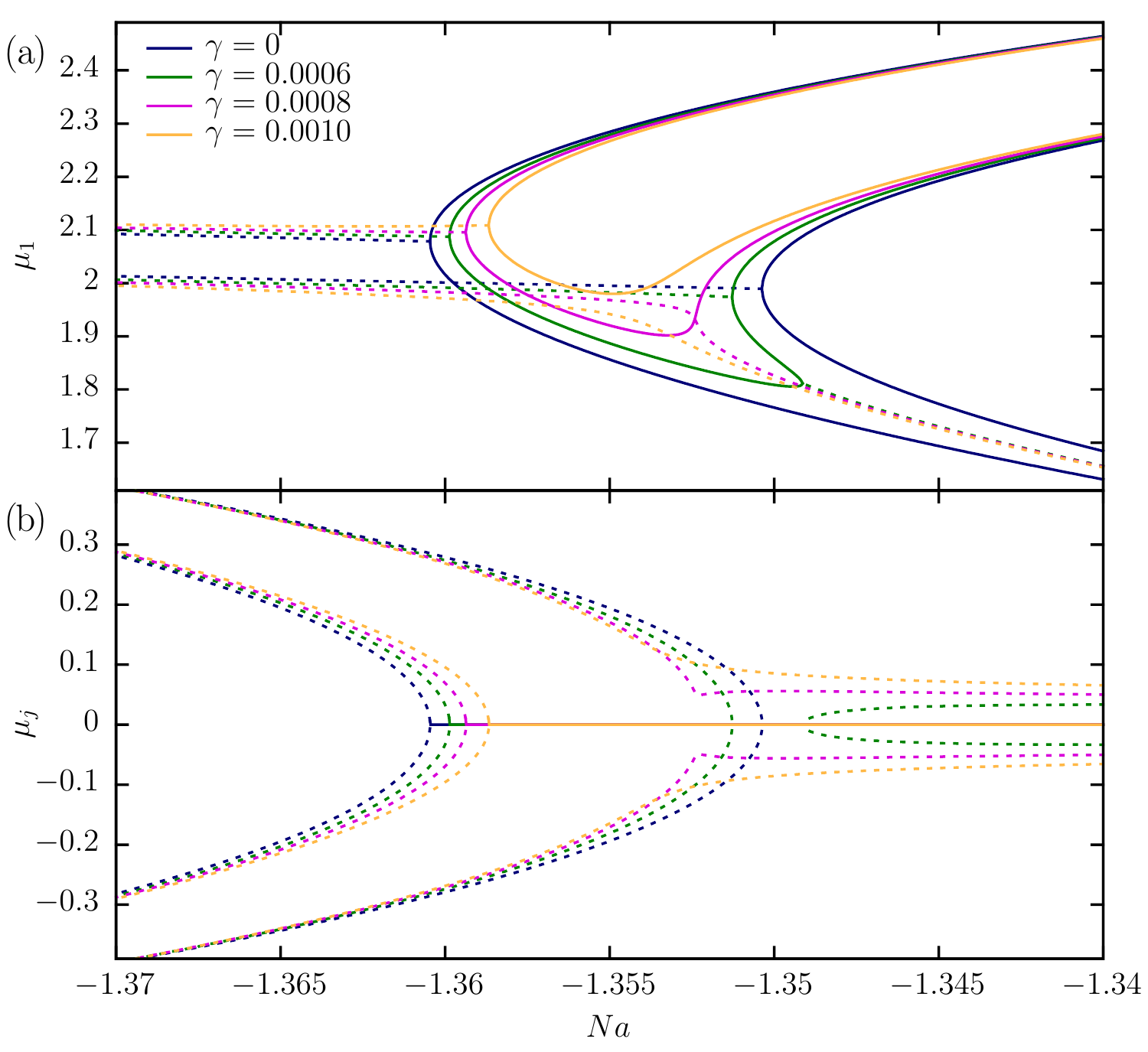}
  \caption{%
    (Color online)
    Bicomplex bifurcation scenario showing the $\mu_1$ and $\mu_\rmj$ component
    of the chemical potential as a function of the nonlinearity parameter $Na$
    for different values of the gain-loss parameter $\gamma$.
    The $\rmi$ and $\rmk$ components vanish for all states discussed, thus
    $\mu^* = \mu$ holds and all states are $\PT$ symmetric.
    The states with purely real chemical potential, i.e.\ $\mu_\rmj = 0$, are
    still present (solid lines).
    In addition a pair of bicomplex states with $\mu = \mu_1 \pm \rmj \mu_\rmj$
    emerges at the tangent bifurcations T1, T2, and T3 (dotted lines).
    For $\gamma > \gamma_\mrm{c}$, the bifurcations T2 and T3 have vanished
    leading to a bicomplex branch that is completely independent from the
    remaining scenario.
    Independent of the parameters $\gamma$ and $Na$ there are four stationary
    solutions.
  }%
  \label{fig:bicomplex_spectrum}
\end{figure}

The bicomplex chemical potential $\mu$ of the stationary solutions is shown in
Fig.~\ref{fig:bicomplex_spectrum}.
For the discussion of the results we again use the components of bicomplex
numbers as introduced in Eq.~\eqref{eq:bicomplex_components} instead of the
representation in the idempotent basis since this renders the interpretation of
the results much clearer.
The first thing to note is that all states have vanishing $\mu_\rmi$ and
$\mu_\rmk$ components.
Therefore $\mu^* = \mu$ holds and all states considered are $\PT$ symmetric.
All states discussed so far are still present in the spectrum and only have a
non-vanishing $\mu_1$ component (solid lines).
However, additional states with $\mu_\rmj \neq 0$ are found (dotted lines).
Since the potential in the Gross-Pitaevskii equation is symmetric under a
complex conjugation with respect to $\rmj$, these states have to appear in
pairs $\mu = \mu_1 + \rmj \mu_\rmj$ and $\mu = \mu_1 - \rmj \mu_\rmj$
\cite{Dast13b}.

First we discuss the case $\gamma < \gamma_\mrm{c}$ where all three tangent
bifurcations are present.
At the bifurcation T1 the states $\ket{\mrm{A}}$ and $\ket{\mrm{D}}$ coalesce
and vanish.
On the other side of the bifurcation two bicomplex states emerge whose chemical
potentials are complex conjugate with respect to $\rmj$.
Also at the bifurcations T2 and T3 two states of the real spectrum vanish and
on the other side of the bifurcations two bicomplex states emerge.
For $\gamma > \gamma_\mrm{c}$ the situation at the bifurcation T1 has not
changed, however, the bifurcations T2 and T3 have vanished and the states
$\ket{\mrm{B}}$ and $\ket{\mrm{D}}$ have merged.
The bicomplex states that emerged from T2 and T3 have also merged and form
independent branches that do not bifurcate with any real branch.

At the tangent bifurcations two eigenvectors and the corresponding eigenvalues
become equal qualifying them as exceptional points of second order
\cite{Heiss12a, Demange12a}.
The bicomplex analysis reveals that at the cusp point, where the bifurcations
T2 and T3 merge, a total of three states, one with real eigenvalue and two with
bicomplex eigenvalues, coalesce.
This is characteristic of an exceptional point of third order which has
already been observed in spectra with similar cusp-like behavior
\cite{Gutohrlein13a, AmShallem14a}.

\begin{figure}
  \centering
  \includegraphics[width=\columnwidth]{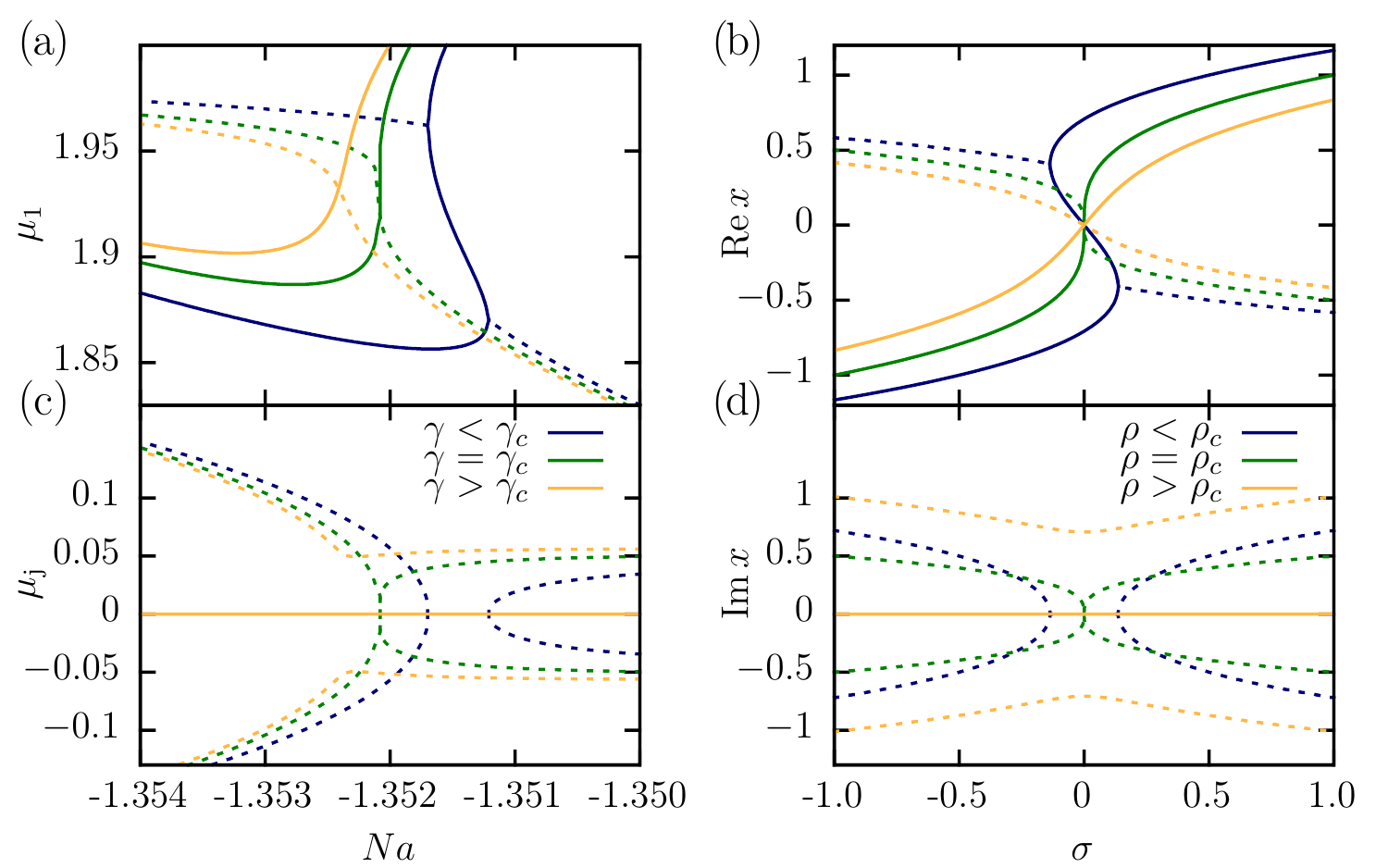}
  \caption{%
    (Color online)
    Comparison of the cusp bifurcation occurring in the eigenvalue spectrum
    (left panels) and in the normal form \eqref{eq:cusp_normalform} (right
    panels).
    For parameter values smaller than the critical value ($\gamma <
    \gamma_\mrm{c}$, $\rho < \rho_\mrm{c}$) the tangent bifurcations T2 and T3
    are separated.
    They coalesce at the critical parameter value ($\gamma = \gamma_\mrm{c}$,
    $\rho = \rho_\mrm{c}$) and vanish for $\gamma > \gamma_\mrm{c}$, $\rho >
    \rho_\mrm{c}$.
    Real branches are drawn as solid lines and (bi)complex branches as dotted
    lines.
  }%
  \label{fig:normalform_spectrum}
\end{figure}

The coalescence of two tangent bifurcations is the characteristic property of a
cusp bifurcation \cite{Poston78a} which is described by the normal form
\begin{equation}
  \dot{x}=x^3 + \rho x - \sigma,
  \label{eq:cusp_normalform}
\end{equation}
whose stationary solutions $\dot{x}=0$ are found by Cardano's method.
The spectrum of the normal form has two tangent bifurcations which vanish at
the critical values $\rho_\mrm{c}=\sigma_\mrm{c}=0$.
In Fig.~\ref{fig:normalform_spectrum}
the normal form is compared with the eigenvalue spectrum of the double-well
potential.
The parameters $\rho$ and $\sigma$ play the role of the gain-loss parameter
$\gamma$ and the nonlinearity parameter $Na$, respectively.
At the tangent bifurcations in the eigenvalue spectrum real values of the
chemical potential $\mu$ turn into bicomplex values with a $\rmj$ component,
$\mu=\mu_1 + \rmj \mu_\rmj$, whereas in the spectrum of the normal form real
values become ordinary complex numbers.
Thus, we have to compare $\mu_1$ with $\real x$ and $\mu_\rmj$ with $\imag x$.
Again the real branches are drawn as solid lines whereas the complex and
bicomplex branches are drawn as dotted lines.

In the regime $\gamma < \gamma_\mrm{c}$, $\rho < \rho_\mrm{c}$ the two tangent
bifurcations are still separated, at $\gamma = \gamma_\mrm{c}$, $\rho =
\rho_\mrm{c}$ the tangent bifurcations coalesce and finally for $\gamma >
\gamma_\mrm{c}$, $\rho > \rho_\mrm{c}$ the bifurcations have vanished.
The normal form captures the qualitative behavior of the bifurcation scenario
found in the eigenvalue spectrum, thus justifying its classification as a cusp
bifurcation.

\section{Conclusion and outlook}
\label{sec:conclusion}
This answers the question raised in \cite{Haag14a} regarding the bifurcation
scenario at strong attractive interactions and, thus, completes the discussion
of the eigenvalue spectrum of the three-dimensional $\PT$-symmetric double-well
potential.
Formulating the TDVP with bicomplex numbers in the idempotent basis has proved
to be useful to analytically continue the non-analytic Gross-Pitaevskii
equation and analyze the bifurcation scenarios in the eigenvalue spectrum.
Using this formalism will help tackling problems with more complicated
$\PT$-symmetric potentials and additional interactions such as the dipolar
interaction which might show an even richer variety of bifurcation scenarios.

\end{document}